\documentstyle[epsfig]{fbssuppl}

\title{Kaon Photoproduction with Form Factors in a Gauge-invariant Approach}

\author{H. Haberzettl\instnr{1}, C. Bennhold\instnr{1}, T. Mart\instnr{1,2}, T. Feuster\instnr{3}}

\instlist{Center for Nuclear Studies, Physics Department, The George Washington University, Washington, DC 20052, U. S. A.
\and Jurusan Fisika, FMIPA, Universitas Indonesia, Depok 16424, Indonesia
\and Institut f\"ur Theoretische Physik, Universit\"at Gie{\ss}en, D-35392 Gie{\ss}en, Germany}

\begin{document}

\maketitle

\begin{abstract}  
The general gauge-invariant photoproduction formalism given by Haberzettl is 
applied to kaon photoproduction off the nucleon at the tree level, with form factors describing composite nucleons. 
We demonstrate that, in contrast to Ohta's gauge-invariance prescription, this formalism
allows electric current contributions to be multiplied by a form factor, i.e., they do not need to
be treated like bare currents. Numerical results show that Haberzettl's gauge procedure, when compared
to Ohta's, leads to 
much improved $\chi^2$ values. Moreover, predictions for the new Bonn SAPHIR data
for $p(\gamma,K^+)\Lambda$ are given.

\end{abstract}


\vspace{2mm}

\begin{figure}[t!]%
\centerline{\psfig{file=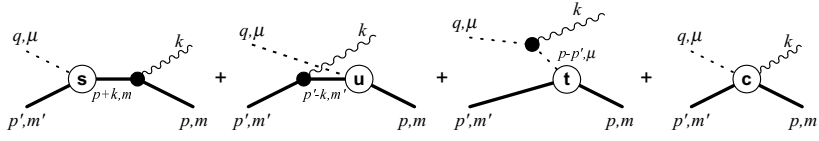,width=0.9\columnwidth,clip=,silent=}} 
\caption[fig1]{\label{tree} Tree-level photoproduction diagrams. Time proceeds
from right to left. 
The form factors $F_s$, $F_u$, and $F_t$ in the text describe
the vertices labeled $s$, $u$, and $t$, respectively, with appropriate
momenta and masses shown for their legs. The right-most diagram
corresponds to the contact current $M_{\text{c}}^\mu$ required to restore gauge invariance.}
\end{figure}

Gauge invariance is one of the central issues
in dynamical descriptions of how photons interact with 
hadronic systems (see Ref.\ \cite{hh97g}, and references therein).
For the simple example of 
$\gamma p \rightarrow n \pi^+$ with pseudoscalar coupling for the $\pi NN$ vertex,
one finds already at the tree level (see Fig.\ \ref{tree}) that the corresponding
amplitude violates gauge invariance if the baryon structure is described by form factors.
This amplitude may be written as \cite{hh98k}
\begin{equation} \epsilon \cdot \widetilde{M}_{fi}=\sum_{j=1}^{4} \widehat{A}_j \overline{u}_n
\left(\epsilon_\mu M_j^\mu \right) u_{\rm p} -2ge\overline{u}_{\text{n}} 
\gamma_5 \epsilon_\mu\bigg[ 
p'^\mu\frac{\widehat{F}-F_s}{s-m^2}
 + q^\mu\frac{\widehat{F}-F_t}{t-\mu^2}\bigg] u_{\text{p}}\;\;, 
\label{mviol}
\end{equation}
with individually gauge-invariant currents,
\begin{eqnarray}
M_1^\mu &=& -\gamma_5 \gamma^\mu \; k \cdot \gamma \;\;, \label{m1}\\
M_2^\mu &=& 2\gamma_5 \left( p^\mu \; k\cdot p' - p'^\mu k\cdot p \; \right)  \;\;, \label{m2}\\
M_3^\mu &=&  \gamma_5 \left( \gamma^\mu \; k\cdot p  - p^\mu  \; k\cdot \gamma  \right)  \;\;, \label{m3}\\
M_4^\mu &=&  \gamma_5 \left( \gamma^\mu \; k\cdot p' - p'^\mu \; k\cdot \gamma  \right)  \;\;, \label{m4}
\end{eqnarray}
with coefficient functions
\begin{eqnarray}
\widehat{A}_1 &=&  \frac{ge}{s-m^2}\left( 1+\kappa_{\rm p}\right) F_s 
       + \frac{ge}{u-m^2}\kappa_n F_u \;\;,\label{ah1}\\
\widehat{A}_2 &=&  \frac{2ge}{(s-m^2)(t-\mu^2)} \widehat{F} \;\;,\label{ah2}\\
\widehat{A}_3 &=&  \frac{ge}{s-m^2}\frac{\kappa_p}{m} F_s \;\;,\label{ah3}\\
\widehat{A}_4 &=&  \frac{ge}{u-m^2}\frac{\kappa_n}{m} F_u \;\;,\label{ah4}
\end{eqnarray}
and a gauge-violating term given by the last term in Eq.\ (\ref{mviol}).

With all external legs on-shell in Fig.\ \ref{tree}, the respective form factors are only functions of
one of the Mandelstam variables, $s$, $u$, or $t$, i.e.,
\begin{eqnarray}
F_s &=& F_s(s) \;=\; f\!\left((p+k)^2,m'^2,\mu^2\right) \;\;, \label{f1}\\
F_u &=& F_u(u) \;=\; f\!\left(m^2,(p'-k)^2,\mu^2\right)\;\;, \label{f2}\\
F_t &=& F_t(t) \;=\; f\!\left(m^2,m'^2,(p-p')^2\right)  \;\;, \label{f3}
\end{eqnarray}
where $f\!(p^2,{p'}^2,q^2)$ is a general $\pi NN$ 
form factor depending on the squared four-momenta of its three hadronic legs.

The function $\widehat{F}$ appearing here 
cancels out in Eq.\ (\ref{mviol}), and hence it is undetermined. Introducing this free function here 
allows us to write
Eq.\ (\ref{mviol}) so that the gauge-invariant limit of having no form factors, {\it viz.}
%
\begin{equation}\label{noff}
\text{Point-like nucleons: \quad}F_s= F_u=F_t=\widehat{F}=1  \; , 
\end{equation}
immediately provides for vanishing of the gauge-violating contribution to the amplitude (\ref{mviol}).

For extended nucleons, and without a detailed dynamical treatment of the compositeness of nucleons \cite{hh97g},
any prescription for restoring gauge invariance amounts 
to introducing an additional
contact current $M_{\rm c}^\mu$ (generically depicted by the fourth diagram in Fig.\ \ref{tree}),
with on-shell matrix elements cancelling exactly the gauge-violating term in Eq.\ (\ref{mviol}).
Apart from purely transverse components, for the present example this contact current is
essentially given by the term in the square brackets of Eq.\ (\ref{mviol}).
Adding this contact contribution to Eq.\ (\ref{mviol}),  one then obtains 
a gauge-invariant amplitude,
\begin{equation} \epsilon \cdot \widehat{M}_{fi}=\sum_{j=1}^{4} \widehat{A}_j \overline{u}_n
\left(\epsilon_\mu M_j^\mu \right) u_p\;\;, 
\label{mgauge}
\end{equation}
which {\it does} depend on $\widehat{F}$ now via $\widehat{A}_2$ of Eq.\ (\ref{ah2}). In other words,
we may use $\widehat{F}$ to distinguish between different choices for repairing gauge invariance.

One of the most popular prescriptions for restoring gauge invariance is due to Ohta \cite{ohta89}.
Using analytic continuation and minimal substitution, 
Ohta finds that the required $\widehat{F}$ is constant,
\begin{equation}  
  \mbox{Ohta: \quad} \widehat{F} =f\!\left(m^2,m'^2,\mu^2\right)= 1\;\;,   
\label{ohtaf}
\end{equation}
determined by the normalization condition for the form factor in the unphysical region where all 
three legs are on-shell. This corresponds precisely to what one obtains for $\widehat{F}$ in
the structureless case (\ref{noff}) and therefore the electric term $\widehat{A}_2$ of Eq.\ (\ref{ah2})
is treated as in the bare case, thus effectively freezing all degrees of freedom arising from the compositeness
of the $\pi NN$ vertex.

The general meson photoproduction theory of Ref.\ \cite{hh97g} provides another, more flexible, way of choosing
$\widehat{F}$. Haberzettl's formalism allows one to take
$\widehat{F}$ as a linear combination
of all form factors appearing in the problem, i.e.,
\begin{equation}
\text{Haberzettl: \quad}
\widehat{F} = a_s F_s(s)+a_u F_u(u)+a_t F_t(t)\;\;,\label{hhb}
\end{equation}
where the coefficients are restricted by $a_s+a_u +a_t=1$ in order to provide the proper limit
for vanishing photon momentum (see Ref.\ \cite{hh98k} for details).

We have tested the relative merits of both prescriptions 
for repairing gauge invariance 
for the kaon photoproduction reactions $\gamma p \rightarrow \Lambda K^+$ and $\gamma p \rightarrow \Sigma^0 K^+$.
In both cases, one can take over Eqs.\ (\ref{ah1})-(\ref{ah4}) and (\ref{mgauge}) 
by replacing the pion by $K^+$ and the
neutron by the respective hyperon. The underlying resonance model we use is the one of Ref.\ \cite{terry}.
For simplicity, we employ here the same hadronic form factor for all resonances, parameterized as
\begin{equation}
f \!\left( p'^2,p^2,q^2 \right) = {\Lambda^4}/\left({\Lambda^4+(  p^2    - m^2)^2
               +( p'^2 - m'^2 )^2 
               +(  q^2 -\mu^2 )^2}\right)\;\;,\label{fgen}
\end{equation}
where $\Lambda$ is some cutoff parameter.

\begin{figure}[t!]
\begin{minipage}[t]{.31\columnwidth}
\epsfig{file=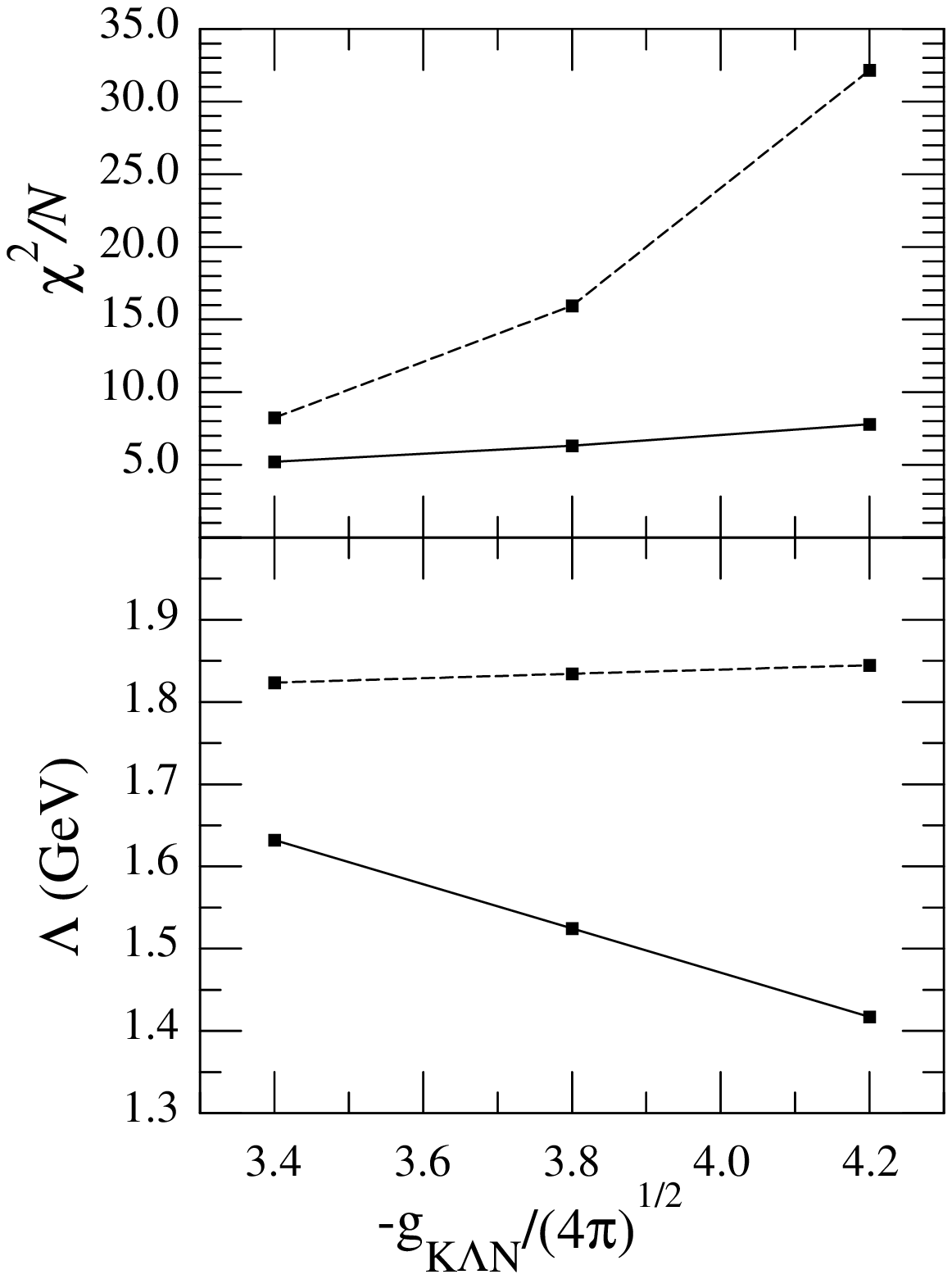,width=\columnwidth,clip=,silent=}
\end{minipage}\hfill\begin{minipage}[t]{.66\columnwidth}
\epsfig{file=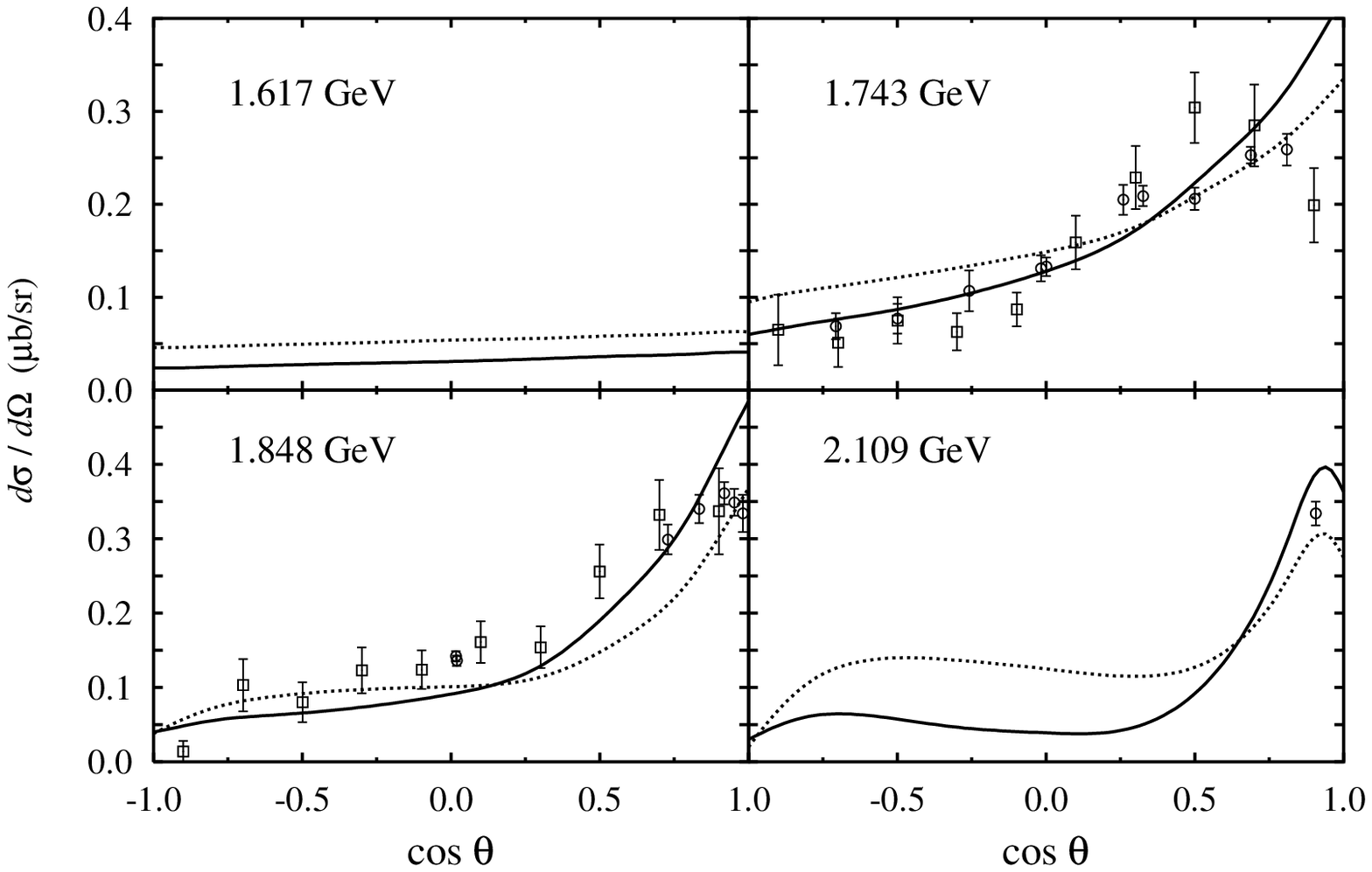,width=\columnwidth,clip=,silent=}
\end{minipage}\\
\begin{minipage}[b]{0.4\columnwidth}
\caption{\label{chi}  $\chi^2/N$ and cutoff parameter $\Lambda$ as functions of
coupling constant  $-g_{K\Lambda N}$ (solid
lines: Eq.\ (\ref{hhb}); dashed lines: Eq.\ (\ref{ohtaf})).}
\end{minipage}\hfill\begin{minipage}[b]{0.56\columnwidth}
\caption{\label{saphir} Differential cross sections for $p(\gamma, K^+)\Lambda$ (solid
lines: Eq.\ (\ref{hhb}); dashed lines: Eq.\ (\ref{ohtaf}); experimental points:
old Bonn SAPHIR data \cite{saphir94}).}
\end{minipage}
\end{figure}

One of the main numerical results is summarized in Fig.\ 2. The upper panel shows 
$\chi^2$ per data point as a function of one of the leading Born coupling constants,
$g_{K \Lambda N} / \sqrt{4 \pi}$, for the two 
different gauge prescriptions by
Ohta and Haberzettl ($g_{K \Lambda N}$ was chosen here because $\chi^2$ shows very little
sensitivity on the other leading coupling constant, $g_{K \Sigma N}$ \cite{hh98k}).  
Clearly, Haberzettl's method provides $\chi^2$ values better than Ohta's by at least a factor of two, 
which, moreover, are almost independent of $g_{K \Lambda N}$, in stark contrast to Ohta's. 
In the fits the form factor cutoff $\Lambda$ was allowed
to vary freely. As is seen in the lower panel of Fig.\ 2, in the case of
Haberzettl's method, the cutoff decreases with
increasing $K \Lambda N$ coupling constant, leaving the magnitude of the
{\it effective} coupling, i.e., coupling
constant times form factor,  roughly constant. 
Since Ohta's method
does not involve form factors for electric contributions [cf.\ Eqs.\ (\ref{ah2}) and (\ref{ohtaf})] no such
compensation is possible there, and as a consequence
the cutoff remains insensitive to the coupling constant (see Ref.\ \cite{hh98k} for more details).

Figure 3 shows differential cross sections for $p(\gamma, K^+)\Lambda$ for four energies
for which new, as yet unpublished, Bonn  SAPHIR data exist. In the figure, we show the old SAPHIR data \cite{saphir94}.
The new data have {\it not} been included in our fit and the curves shown in Fig.\ 3 are, therefore, 
{\it predictions}. As will be seen when they become available publicly, 
the new data clearly favor the gauge-invariance prescription by Haberzettl.

Our overall conclusion from the present findings is that Ohta's approach seems too restrictive
to account for the full hadronic structure while properly maintaining gauge invariance,
whereas the method put forward in Ref.\ \cite{hh97g} seems well capable of providing this facility.

This work was supported in part by Grant No.\ DE-FG02-95ER40907 of the U.S. Department of 
Energy.

\makeatletter \if@amssymbols%
\clearpage 
\else\relax\fi\makeatother

\SaveFinalPage
\end{document}